# Molecular Aspect of Annelid Neuroendocrine system

Michel Salzet

Laboratoire de Neuroimmunologie des Annélides, UMR CNRS 8017, SN3, Université des Sciences et Technologies de Lille, 59655 Villeneuve d'Ascq Cedex, France.

Tel : + 33 3 2033 7277 ; Fax : + 33 3 2043 4054, email : michel.salzet@univ-lille1.fr;
http://www.univ-lille1.fr/lea

**Running Title** : Annelid's nervous system

## ABSTRACT
Hormonal processes along with enzymatic processing similar to that found in vertebrates occur in annelids. Amino acid sequence determination of annelids precursor gene products reveals the presence of the respective peptides that exhibit high sequence identity to their mammalian counterparts. Furthermore, these neuropeptides exert similar physiological function in annelids than the ones found in vertebrates. In this respect, the high conservation in course of evolution of these molecules families reflects their importance. Nevertheless, some specific neuropeptides to annelids or invertebrates have also been in these animals.

## Table of Contents

**Introduction**

Annelids neural tissues do not contain anatomical correlates of hypothalamus or pituitary. However, they possess localized ganglionic regions rich in mammalian-like neuroendocrine signaling molecules [1]. These molecules appear free in the animals' hemolymph, demonstrating distant signaling *via* several target tissues, including immune cells [1, 2]. Thus, the fact that a classical closed conduit system to carry signaling molecules does not exist in Annelids should not detract from an endocrine presence because the baseline functioning of the system depends on distance between the origin of a signal molecule and its target tissue/receptor.

**1. Neuroendocrine peptide families in Annelids similar to Vertebrates ones**

Among the 40 neuropeptides isolated, so far, in annelids (Tables 1 and 2), the most part already sequenced are related to the ones previously isolated in vertebrates [1-5]. More recently some EST databanks of annelids have been carried out by several groups and have opened the door of the discovering of novel neuropeptide families like the angiotensins, the oxytocin/vasopressin, the myotropic peptides, and the opioid families. With the exception to the myotropic peptides, which are more specific to the annelids, the other molecules families have also been identified in vertebrates in which they play crucial roles as neurohormones. These data further strengthen the existence of a neuroendocrine system in the annelids.

*1.1. Angiotensin-like peptides*

Biochemical identification of a "central" angiotensin II (AII)-like peptide in the leech *Erpobdella octoculata* was demonstrated and found to be amidated. [6-9]. This constituted the first characterization of an angiotensin-like peptide in an invertebrate, demonstrating its conservation during the course of evolution. An identification of the proteins immunoreactive to anti-AII was found both in brain extracts and *in vitro* translated brain RNA products [6]. The pro-AII precursor detected in the brain extracts possesses a *ca* 19 kDa molecular mass and is also a "multiple hormone precursor" as it is also recognized by two other antisera : a polyclonal γ-MSH and a specific monoclonal antibody directed against leech neurons, Tt159 [10]. Furthermore, we found in leeches a *ca.* 11-kDa peptide with a sequence of DRVYIHPFHLLXWG, which exhibits a 78.5% sequence identity to the N-terminus of the angiotensinogen and a 100% sequence identity to AI [7].

Biosynthesis study of leech AII revealed the existence of a renin- [11] and angiotensin-converting like enzymes (ACE)-like enzymes implicated in its catabolism [13]. Leech ACE has been cloned and an approximative 2 kb cDNA has been predicted to encode a 616-amino-acid soluble enzyme containing a single active site, named *Tt*ACE (*T. tessulatum* ACE) [14]. Surprisingly, its primary sequence shows greater similarity to vertebrates than to invertebrates. Stable *in vitro* expression of *Tt*ACE in transfected Chinese-hamster ovary cells revealed that the leech enzyme is a functional metalloprotease. As in mammals, this 79 kDa glycosylated enzyme functions as a dipeptidyl carboxypeptidase capable of hydrolysing angiotensin I to angiotensin II. However, a weak chloride inhibitory effect and acetylated *N*acetyl-SDKP (Ac SDAcKP) hydrolysis reveal that *Tt*ACE activity resembles that of the N-domain of mammalian ACE [14]. *In situ* hybridization shows that its cellular distribution is restricted to epithelial midgut cells [14]. EST screening reveals that both membrane and soluble somatic forms of ACE are present in leeches. In the polychaete *Neanthes virens* a Dipeptidyl carboxypeptidase resembling mammalian ACE has also been demonstrated [15, 16].The purified enzyme was homogeneous by SDS-PAGE, with a molecular mass of 71 kDa by SDS-PAGE and 69 kDa by gel filtration, indicating that it is monomeric like the leech *Tt*ACE [14]. The isoelectric point was 4.5 and optimum pH for the activity was 8.0. It showed a specific activity of 466.8 U/mg [16].

Experiments conducted on the biological activity of the AIIamide established that this peptide is involved in the control of leeches' hydric balance exerting a diuretic effect [17]. Similarly, in the Polychaeta *Nereis diversicolor*, injections of polyclonal antisera against AII provoked a partial inhibition of the increase in body weight in animals exposed to hypo-osmotic medium. In a subsequent test, injections of synthetic AII-amide and, to a lesser extent AII enhanced the increase in body weight and, therefore, strengthened the importance of these peptides in the neuroendocrine control of *Nereis* osmoregulation [18]. In clam worm *Perinereis* sp, angiotensin III as well as angiotensin II enhanced an increase in body weight under a hypo-osmotic condition and suppressed a decrease in body weight under a hyper-osmotic condition. When clam worms were treated with tetrachloroaurate (III) after angiotensin-treatment, these enhancing and suppressive effects of the angiotensins under hypo- and hyper-osmotic conditions were inhibited. In contrast, when clam worms were pretreated with tetrachloroaurate (III) before angiotensin-treatment, these effects of angiotensins were not inhibited. Since tetrachloroaurate (III) is a representative blocker of aquaporins, these results indicate that angiotensin III as well as angiotensin II regulates water flow through aquaporins in clam worms [19, 20]

In leeches, the fact that AIIamide injections at different doses in *T. tessulatum* suggests the existence of two different types of receptors, one at high and the other one at low affinities towards AIIamide, we focused our interest on the identification of the leech AIIamide receptors. Binding experiments on *T. tessulatum* brain membranes with mono [$^{125}$I]AIIamide reveals 70% specific binding and a $IC_{50}$ of 10 nM [21]. In addition, biochemical studies using commercial anti-AT1 receptor reveal the existence of a specific protein at a molecular weight of 140 kDa [21] which is confirmed by EST analysis where AT1 and AT2 like receptors have been detected with 27 and 32% sequence identity, respectively. Immunocytochemical studies performed at the level of the brain confirmed the presence of labeling in neurons and glial cells to anti-AII, anti-leech renin, anti-leech ACE and anti-AT1 [21]. Leech coelomocytes are also immunoreactive to polyclonal antisera raised against the *T. tessulatum* ACE and leech brain angiotensin II (AII) and a commercial anti-AT1 receptor. Biochemically, renin, ACE and AT1-like receptor were identified in the leech immune cells [21]. Moreover, leech AII ($10^{-6}$ M) alone does not initiate nitric oxide (NO) release in invertebrate immunocytes but does only after pre-exposing the cells to IL-1 (15.9+/-2.6 nM; P<0.005 vs. 1.1 nM when AII is added alone). Similar results were obtained with human leukocytes (14.5+/-2.7 nM; P<0.005 IL-1+AII vs. 0.9 nM when AII is added alone). Immunocytochemical studies performed at the structural and ultrastructural levels confirmed the presence in same immune cells all the molecules of the renin-angiotensin system (RAS) in leeches as epitopes to IL-1-like protein and IL-1-like receptor. These data report a co-action between cytokines like substances and neuropeptides in an immune process and the involvement of the RAS in modulation of the immune response [21].

Complementary to these findings, evidences of natural ACE inhibitors have been demonstrated in leeches [22]. In fact, the leech osmoregulator Factor (IPEPYVWD, see the LORF section), a neuropeptide found in both central nervous system and sex ganglia of leeches is involved in water retention control through inhibition of ACE with an $IC_{50}$ of 19.8 µM for rabbit ACE. Its cleavage product the IPEP is a better competitive inhibitor with an $IC_{50}$ of 11.5µM. Competition assay using p-[$^{3}$H]benzoylglycylglycylglycine and insect ACE established that LORF and IPEP are natural inhibitors for invertebrate ACE. 54% of insect ACE activity is inhibited with 50 µM d'IPEP and 35% with 25 µM de LORF like [22].

Taken together, these data strongly suggest the existence of a complete RAS in annelids which was conserved in course of evolution.

*1.2. Oxytocin/vasopressin peptides*

The peptides of the vasopressin/oxytocin family have been discovered throughout the animal kingdom. They are alike, sharing at least five of nine residues and a disulfide-linked ring structure, which puts severe constraints on conformational flexibility. In vertebrates, gene duplication gave rise to two distinct families *i.e.* the vasopressin (VP) one and the oxytocin (OT) one (Table 3). The differential binding of VP and OT to their respective receptors is largely due to the amino acid residue in position 8 *i.e.* a basic one in VP-related peptides and a neutral one in OT-related peptides. In the leech *Erpobdella octoculata*, a VP related peptide; the lysine-conopressin (CFIRNCPKG) has been isolated [23]. However, EST screening in *Hirudo medicnalis* has revealed a peptide related to lysine-conopressin but belonging to the OT family (CFIRNCPLG), named Hirudotocin (Salzet, Unpublished data). In Oligochaeta, a peptide related to this family (CFVRNCPTG), the annetocin has been discovered [24]. The proannetocin showed 37.4-45.8% amino acid homology to other prohormones. In the neurophysin domain, 14 cysteines and amino acid residues essential for association of a neurophysin with a vasopressin/oxytocin superfamily peptide were conserved, suggesting that the Eisenia neurophysin can bind to annetocin [25]. Furthermore, in situ hybridization experiments demonstrated that the annetocin gene is expressed exclusively in neurons of the central nervous system predicted to be involved in regulation of reproductive behavior [25]. Similar results are obtained for the Hirudotocin (Salzet, unpublished data).

These peptides act on osmoregulation *via* nephridia [23-25]. Lysine-conopressin inhibits the $Na^+$ amiloride dependent transitory current before to highly stimulate it on *Hirudo medicinalis* stomach or tegument preparation [17]. Moreover, lysine-conopressin induces egg laying in earthworm like OT does in vertebrates [26]. This confirms that through the course of evolution the OT/VP peptides family has conserved its function on both osmoregulation and reproduction

Annetocin receptor (AnR) has been cloned [4]. The deduced precursor displays high sequence similarity with OT/VP receptors. Genomic analysis of the AnR gene revealed that the intron-inserted position is conserved between the AnR gene and the mammalian OT/VP receptor genes (Fig. 1). These results indicate that AnR and mammalian OT/VP receptors share a common ancestor gene. Administration of annetocin to the AnR expressed in Xenopus oocytes induced a calcium-dependent signal transduction. Reverse transcriptase-PCR analysis and in situ hybridization showed that the AnR gene is expressed specifically in the nephridia located in the clitellum region, although the nephridia are distributed throughout the worm body. This result suggests that annetocin induces egg-laying behaviour through its action on the nephridia. This is the first description concerning the functional correlation between an invertebrate OT/VP-related peptide and egg-laying behaviour [4]. In leeches, a 1.7 kb cDNA for an AVP-related receptor has been cloned from the leech *Theromyzon tessulatum* [27]. The open reading frame encodes a 435-amino acid transmembrane protein that displays seven segments of hydrophobic amino acids, typical of G-protein-coupled receptors. The overall predicted protein exhibits about 30% amino-acid identities to other invertebrate, as well as vertebrate, AVP/OT receptor family members, and displays conserved characteristic features belonging to the AVP/OT receptor superfamily. RT-PCR expression experiments showed that mRNA is expressed in the genital tract, the ovary and the brain. The receptor expression is stage specific, showing a weak expression after the two first blood meals, increasing dramatically after the last blood meal during the period of sexual maturation and disappearing after egg laying. Thus, the leech AVP-related receptor may mediate reproductive functions. When expressed in COS-7 cells, the receptor binds ligands with the following rank order of potency: AVP= Arg-vasotocin >Arg-conopressin >mesotocin = OT = Lys-conopressin=isotocin>annetocin. This shows an AVP-like pharmacological profile. The transfected receptor mediates AVP-induced accumulation of inositol phosphates, indicating that the leech AVP-related receptor is functional [27].

Taken together, these data demonstrate the characterization of AVP/OT superfamily receptor in annelids, which are considered the most distant group of coelomate metazoans possessing a functional AVP/OT-related endocrine system [4 ,27].

*1.3. Opioids*

Enkephalins have been isolated in Annelids [28] and its precursor the proenkephalin shows sequence similarity with amphibian proenkephalin (26.2 %) (29, 30). The proenkephalin contains Met- and Leu-enkephalins in a ratio of ½ and Met-enkephalin-Arg-Gly-Leu and Met-enkephalin-Arg-Phe flanked by dibasic amino acid residues, which are targets for proteolytic enzymes such like prohormone convertases present in leeches ([29], Fig. 1). Specific receptors for these peptides have been characterized in neural and in immune systems [29]. Enkephalin derived peptides seem to be implicated in innate immune response (31). In fact, lipopolysaccharides (LPS) injection into the coelomic fluid of the leech *Theromyzon tessulatum* stimulates release of proenkephalin A (PEA)-derived peptides as determined by immunoprecipitation and Western blot analyses [32]. This release occurs in the first 15 min after LPS exposure and yields a 5.3-kDa peptide fragment corresponding to the C-terminal part of the precursor. This fragment is then cleaved to free an antibacterial peptide related to mammals arginine phenylalanine extended enkelytin: the peptide B. These PEA processing peptides were characterized using a combination of techniques including reversed-phase HPLC, microsequencing and mass spectrometry. The isolated invertebrate peptide B presents a high sequence homology with the bovine's and the same activity against Gram+bacteria. Titrations revealed the simultaneous appearance of Methionine-enkephalin (ME) and peptide B in invertebrates after stimulation by LPS (in a dose-dependent manner), surgical trauma or electrical stimulations to neural tissues of the mussel [32]. Furthermore, peptide B processing in vitro yields Methionine-enkephalin arginine phenylalanine (MERF), which exhibits via the delta receptors, immunocyte excitatory properties, *i.e.*, movement and conformational changes, but no antibacterial activity. We surmise that this unified response to the various stimuli is a survival strategy for organism by providing immediate antibacterial activity and immunocyte stimulation, thereby reducing any immune latency period needed for an adequate immune response [30, 32-34].

Besides enkephalins, two other opioids derived from two other opioid precursors have been characterized, α-Neo-endorphin [35] and γ-MSH [36]. Characterization of the entire pro-dynorphin opioid precursor revealed that it exhibits a 28.8% sequence identity to rat, and 22% to the human and pig [37]. Although the α Neo endorphin is identical to the one found in vertebrates, the dynorphins are slightly shorter. A POMC-like molecule has also been demonstrated in leeches [38]. Of the six peptides, three showed high sequence similarity to their vertebrate counterparts, namely, met-enkephalin, α-MSH and ACTH (100, 84.6 and 70% respectively) whereas γ-MSH, β-endorphin and γ-LPH exhibited only 45, 20 and 10% sequence identity. No dibasic amino acid residues were found at the C-terminus of the γ- and β-MSH peptides suggesting that they are not produced in the leech or that they could be synthesized via an alternate biosynthetic pathway. In contrast, the leech α-MSH was flanked at its C-terminus by the Gly-Arg-Lys amidation signal. Circulating levels of ACTH and MSH were 10 and 1 fmol/microl hemolymph, respectively. Morphine, in a dose-dependent manner, increased the levels of both peptides threefold; this effect was blocked by naloxone treatment. Similar results were found with the anandamide. Leech ACTH was processed to MSH by the enzymes neutral endopeptidase (24.11) and angiotensin-converting enzyme. Leech α-MSH had the same activity as authentic alpha-MSH in two bioasssay systems [38].

Taken together, the results from leeches now demonstrate, those opioid precursors and many of the derived bioactive peptides, *i.e.*, α-MSH and ACTH, which are important in mammalian neuroendocrine signaling, are present in invertebrates. This adds to the growing body of evidence that a neuroendocrine apparatus is also present in simple animals.

## 2. Neuroendocrine peptide families in Annelids similar to Molluscs ones

Annelids and Molluscs are both lophochotrozoans and in this context several families of peptides have been well conserved in these two groups. The most represented one is the Rxamides family. However other molecules families have been detected in leech and lymnaea or Aplysia but are also now discovered in vertebrates like the RFamide peptide family [7].

### 2.1. RXamide Peptides

Several related RXmides have been found in annelids [39-46]. A myomodulin-like peptide, GMGALRLamide, has been purified and sequenced from the medicinal leech nerve cords (45). Myomodulin-like immunoreactivity has recently been found to be present in a set of leech neurons, including Leydig neurons [45, 46]. The glial responses to Leydig neuron stimulation persisted in high-divalent cation saline, when polysynaptic pathways are suppressed, indicating that the effects on the glial cell were direct [42]. The glial responses to myomodulin A application persisted in high-$Mg^{2+}$/low-$Ca^{2+}$ saline, when chemical synaptic transmission is suppressed, indicating a direct effect of myomodulin A on the glial membrane (42). The glial hyperpolarization evoked by myomodulin A was dose dependent ($EC_{50}$ = 50 nM) and accompanied by a membrane conductance increase of approximately 25%. Ion substitution experiments indicated that myomodulin A triggered a $Ca^{2+}$-independent $K^+$ conductance (42). Moreover, synthetic leech myomodulin-like peptide [46] showed identical neuronal modulation effect on the giant leech Retzius cell comparable to that by the synthetic Aplysia myomodulin APMGMLRLamide [47-49]. This neural and muscular modulation has been shown to be important for shaping and modifying behavior. Experiments focused on the Retzius cell (R) revealed that the myomodulin-like peptide increased the excitability of the R cell such that the cell fires more action potentials with a shorter latency to the first action potential. This effect is mediated by the activation of a Na+-mediated inward current near the cell resting membrane potential [44]. EST library scanning reveals the presence of a myomodulin precursor sharing several peptides of this family bracketed by di-basic residues like GVSFLRMG, SLDMLRMG, AVSMLRMG, AVSLLRMG (Salzet, unpublished data) and presenting 57% sequence homology with Lymnaea or Aplysia myomodulin precursor (49, 50) (Fig. 2).

In polychaete annelids a heptapeptide, AMGMLRMamide, termed Pev-myomodulin, was isolated from *Perinereis vancaurica* using the esophagus of the animal as the bioassay system. The sequence of the annelid peptide is highly homologous with those of the myomodulin-CARP-family peptides found in molluscs. The annelid peptide is regarded as a member of the myomodulin-CARP family, though all the molluscan peptides have a Leu-NH$_2$ at their C-termini. The annelid peptide showed a potent contractile action on the esophagus of the annelid. The peptide may be an excitatory neuromediator involved in the regulation of the esophagus. Among various myomodulin-CARP-family peptides and their analogues, the annelid peptide showed the most potent contractile action on the esophagus [66]. Replacement of the C-terminal Met-NH2 of the annelid peptide with a Leu-NH$_2$ decreased its contractile potency, while replacement of the C-terminal Leu-NH$_2$ of myomodulin and CARP with a Met-NH2 increased their potency. The C-terminal Met-NH$_2$ of the annelid peptide seems to be important, but not essential, for exhibiting its contractile activity on the esophagus [66].

*2.2 RF-amide peptides*

FRaP family is present in the polychaetes *Nereis virens* and *Nereis diversicolor* [51-53]. In this last species, two other RFamide peptides (FM(O)RFamide and FTRFamide) have been isolated [51-53]. Pharmacological data suggest that RFamide peptides are involved in the control of heartbeat and body wall tone in the polychaete *Sabellastarte magnifica* [54, 55] and in the oligochaete *Eisenia foetida* [56].

In the eathworm *Eisenia fetida*, FMRF-like peptides are co-localized with serotonin, suggesting a role as neuromodulators influencing serotoninergic neurons [56, 57]. In Lumbricus *terrestris*, FMRF-like peptides seem to be involved in both central integratory processes, neuromuscular regulation and sensory processes [56, 57].

In Hirudinae, anti-FMRFamide immunoreactivity is found in cell bodies and neuronal processes of the central nervous system [58-63]. In the segmental ganglia of the ventral nerve cord, this immunoreactivity is localized in heart excitatory (HE) motor neurons, heart accessory (HA) modulatory neurons and several motor neurons innervating the longitudinal and medio-dorso ventral muscles [64-66]. Among the 21 segmental ganglia (SG1-SG21) of the ventral nerve cord of leeches, SG5 and SG6 that innervate the sexual organs are designated as sex SG. These sex ganglia contain, as compared to the non-sex ganglia, an additional population of neurons immunostained with anti-FMRFamide in *Hirudo medicinalis* [67]. Furthermore, two RFamide peptides (FMRFamide and FLRFamide) were identified in *Hirudo medicinalis* [60, 68].

These peptides increase the strength and accelerate the rate of myogenic contractions as well as inducing myogenic contractions in quiescent hearts [65, 66]. Besides these tetrapeptides, we characterized an extended form of FLRFamide, the GDPFLRFamide from sex ganglia extracts of *E. octoculata* [60]. In *T. tessulatum* presence of RFamide peptides in neurosecretory granules in fibers of the neurohaemal area suggests that at least one of the characterized peptides is secreted into the dorsal vessel. The brain could exert a neuroendocrine control of certain functions via RFamide peptides. Taking into account a previous study showing a loss of weight of *T tessulatum* after a GDPFLRFamide injection and an increase of weight after a FMRFamide injection [60], we surmise that GDPFLRFamide may act as a diuretic hormone and FMRFamide as an anti-diuretic hormone. Electrophysiological experiments confirmed our speculation [17]. The anti-diuretic effect of FMRFamide seemed not due to a direct action on the caecal epithelium. Nevertheless, the control of the hydric balance might be also exerted directly on the nephridia. Indeed, Wenning *et al*. demonstrated that the nephridial nerve cells, which innervate the nephridia and contact the urine forming cells, contain RFamide peptide(s) in *H .medicinalis* [61, 69]. Furthermore, these authors showed that FMRFamide leads the hyperpolarization and decreases the rate of firing of the nephridial nerve cells, suggesting autoregulation of peptide release.

*2.3. Leech egg-laying hormone*

In leeches, egg-laying may be under the control of a leech egg laying hormone (L-ELH) [70]. In *Eisenia fetida*, although that the OT-VP related peptide, annetocin, is known to potentate the pulsatory contractions in bladder-shaking movement of the nephridia, indicating an involved of osmoregulation though nephridial function, this peptide is also implicated in egg laying behaviors [26]. In fact, annetocin, induced a series of egg-laying-related behaviors in the earthworms. These stereotyped behaviors consisted of well-defined rotatory movements, characteristic body-shape changes, and mucous secretion from the clitellum [26]. Each of these behaviors is known to be associated with formation of the cocoon in which eggs are deposited. In fact, some of the earthworms injected with annetocin (> 5 nmol) laid eggs. Such egg-laying-related behaviors except for oviposition were also induced by oxytocin, but not by Arg-vasopressin. Furthermore, annetocin also induced these egg-laying-like behaviors in the leech *Whitmania pingra*, but not in the polychaete *Perinereis vancaurica*. These results suggest that annetocin plays some key role in triggering stereotyped egg-laying behaviors in terrestrial or fresh-water annelids that have the clitella [26].

**3. Are some Neuroendocrine Signaling Molecules specific to Annelids?**

The history of neurobiology demonstrates the significance of the invertebrate nervous system as a valuable model. The giant axon of the squid and crayfish neuromuscular junction stands out in this regard. This field of scientific endeavor also stands out for its demonstration of the conservation of signaling molecules and their functions during evolution. Probably the first peptide found in invertebrates that later was also found in mammals is the *Hydra* head activator peptide [71-73]. In *Hydra* this peptide modulates morphogenesis, cellular growth and differentiation. In mammals, this peptide sequence proved to be unique in that it differed from that of any known peptide. Furthermore, the investigators surmise that this peptide may have the same functions in mammals. We propose here to extend this several families specifically found in annelids and not yet found in mammals.

*3.1. Leech osmoregulator Factor*

In early reports we found that LORF (IPEPYVWD) is a peptidergic-signaling molecule involved in osmoregulation [74]. Furthermore, electrophysiological experiments conducted in the leech *H. medicinalis* revealed an inhibition of the efficacy of $Na^+$ conductance in leech skin [17]. Immunocytochemical studies with an antiserum against synthetic LORF found a great amount of positive immunoreactive neurons in all ganglia. This material was present in a single type of electron-dense secretory granules of a size of 80-100 nm [32]. However, we recently showed its presence in rat tissues, including discrete brain areas (75), which demonstrate that this peptide, originally isolated from the leech, is also present in mammals. Furthermore, in both the leech central nervous system (CNS) and the rat brain, e.g., hypothalamus, LORF is coupled to nitric oxide (NO) release [75]. It is also capable of stimulating NO release from human saphenous vein fragments [75]. These results showed that LORF is not specific to annelids but they also demonstrated that Leech represents a suitable model to isolate novel peptides also conserved in vertebrates.

*3.2. GGNG Peptides*

The most specific peptides presently isolated in annelids are the ones related to the GGNG, excitatory peptides [76-80] (Fig. 3). In leeches, these peptides elicit muscular contractions of isolated preparation of penis and intestine, suggesting that they may play a role in reproduction behavior like the PLGWamide in mollusks [76-80]. In earthworm, they elicit gut and esophagus muscular contractions. No peptides homologous to GGNG peptides have been isolated so far in any living organisms.

### 4. Are Neuropeptide Processing Enzymes Present in Invertebrates?

The biologically active neuropeptides are produced via the proteolytic processing of precursor molecules. Among the enzymes involved in neuropeptide precursors processing, the serine proteases belonging to the subtilisin/kexin family and known as the proprotein convertases (PCs) are playing, at least in mammals, a crucial role. In mammals, this family includes, so far, seven members (PC1, PC2, furin, PC4, PACE4, PC5 and PC7) which are differentially involved in the processing, generally at single or dibasic residues, of a large number of protein precursors including neuropeptides, growth factors, receptors, enzymes and viral envelope glycoproteins [81]. Between these enzymes, only PC1 and PC2, which are abundantly expressed in neurons and endocrine cells, play an important role in the neuropeptide precursors processing. Although these latter enzymes have been cloned in numerous species including mammals, amphibians, molluscans, insects and nematods, they have not yet been characterized in annelids. EST screening confirms the presence of prohormone convertases in leech brain sharing at least 65% sequence identity.

### 7. Conclusion

In conclusion, given the wealth of information now emerging on these mammalian-like neuroendocrine processes found in invertebrates, it would appear that this system, in all probability, originated in "simple" animals. Therefore, aside from its historical origin, it may be more appropriate to speak of the mammalian neuroendocrine system as Annelid-like.

### Acknowledgments

This work was in part supported by the Centre National de Recherche Scientifique (CNRS) and the MNERT.

**Reference**
1. Lefebvre, C., and Salzet, M. (2003) *Curr Pharm Des* 9, 149-158
2. Chopin, V., Salzet, M., Baert, J., Vandenbulcke, F., Sautiere, P. E., Kerckaert, J. P., and Malecha, J. (2000) *J Biol Chem* 275, 32701-32707
3. Wang, W. Z., Emes, R. D., Christoffers, K., Verrall, J., and Blackshaw, S. E. (2005) *Cell Mol Neurobiol* 25, 427-440
4. Kawada, T., Kanda, A., Minakata, H., Matsushima, O., and Satake, H. (2004) *Biochem J* 382, 231-237
5. Voronezhskaya, E. E., Tsitrin, E. B., and Nezlin, L. P. (2003) *J Comp Neurol* 455, 299-309
6. Salzet, M., Bulet, P., Wattez, C., Verger-Bocquet, M., and Malecha, J. (1995) *J Biol Chem* 270, 1575-1582
7. Laurent, V., Bulet, P., and Salzet, M. (1995) *Neurosci Lett* 190, 175-178
8. Salzet, M., Wattez, C., Baert, J. L., and Malecha, J. (1993) *Brain Res* 631, 247-255
9. Salzet, M., Verger-Bocquet, M., Wattez, C., and Malecha, J. (1992) *Comp Biochem Physiol A* 101, 83-90
10. Chopin, V., Bilfinger, T. V., Stefano, G. B., Matias, I., and Salzet, M. (1997) *Eur J*

*Biochem* 249, 733-738
11. Laurent, V., and Salzet, M. (1995) *Peptides* 16, 1351-1358
12. Laurent, V., and Salzet, M. (1996) *Peptides* 17, 737-745
13. Laurent, V., and Salzet, M. (1996) *FEBS Lett* 384, 123-127
14. Riviere, G., Michaud, A., Deloffre, L., Vandenbulcke, F., Levoye, A., Breton, C., Corvol, P., Salzet, M., and Vieau, D. (2004) *Biochem J* 382, 565-573
15. Kawamura, T., Oda, T., and Muramatsu, T. (2000) *Comp Biochem Physiol B Biochem Mol Biol* 126, 29-37
16. Kawamura, T., Kikuno, K., Oda, T., and Muramatsu, T. (2000) *Biosci Biotechnol Biochem* 64, 2193-2200
17. Milde, H., Weber, W. M., Salzet, M., and Clauss, W. (2001) *J Exp Biol* 204, 1509-1517
18. Fewou, J., and Dhainaut-Courtois, N. (1995) *Biol Cell* 85, 21-33
19. Satou, R., Nakagawa, T., Ido, H., Tomomatsu, M., Suzuki, F., and Nakamura, Y. (2005) *Biosci Biotechnol Biochem* 69, 1221-1225
20. Satou, R., Nakagawa, T., Ido, H., Tomomatsu, M., Suzuki, F., and Nakamura, Y. (2005) *Peptides* 26, 2452-2457
21. Salzet, M., and Verger-Bocquet, M. (2001) *Brain Res Mol Brain Res* 94, 137-147
22. Deloffre, L., Sautiere, P. E., Huybrechts, R., Hens, K., Vieau, D., and Salzet, M. (2004) *Eur J Biochem* 271, 2101-2106
23. Salzet, M., Bulet, P., Van Dorsselaer, A., and Malecha, J. (1993) *Eur J Biochem* 217, 897-903
24. Oumi, T., Ukena, K., Matsushima, O., Ikeda, T., Fujita, T., Minakata, H., and Nomoto, K. (1994) *Biochem Biophys Res Commun* 198, 393-399
25. Satake, H., Takuwa, K., Minataka, H., Matsushima, O. (1999) J. Biol. Chem. 274(9) 5605-5611.
26. Oumi, T., Ukena, K., Matsushima, O., Ikeda, T., Fujita, T., Minakata, H., and Nomoto, K. (1996) *J Exp Zool* 276, 151-156
27. Levoye, A., Mouillac, B., Riviere, G., Vieau, D., Salzet, M., and Breton, C. (2005) Theromyzon tessulatum. *J Endocrinol* 184, 277-289
28. Salzet, M., Bulet, P., Verger-Bocquet, M., and Malecha, J. (1995) *FEBS Lett* 357, 187-191
29. Salzet, M., and Stefano, G. B. (1997) *Brain Res* 768, 224-232
30. Fimiani, C., Arcuri, E., Santoni, A., Rialas, C. M., Bilfinger, T. V., Peter, D., Salzet, B., and Stefano, G. B. (1999) *Cancer Lett* 146, 45-51
31. Salzet, M. (2001) *Neuro Endocrinol Lett* 22, 467-474
32. Tasiemski, A., Verger-Bocquet, M., Cadet, M., Goumon, Y., Metz-Boutigue, M. H., Aunis, D., Stefano, G. B., and Salzet, M. (2000) *Brain Res Mol Brain Res* 76, 237-252
33. Salzet, M. (2000) *Brain Res Rev* 34, 69-79
34. Bilfinger, T. V., Salzet, M., Fimiani, C., Deutsch, D. G., Tramu, G., and Stefano, G. B. (1998) *Int J Cardiol* 64 Suppl 1, S15-22
35. Salzet, M., Verger-Bocquet, M., Bulet, P., Beauvillain, J. C., and Malecha, J. (1996) *J Biol Chem* 271, 13191-13196
36. Salzet, M., Wattez, C., Bulet, P., and Malecha, J. (1994) *FEBS Lett* 348, 102-106
37. Salzet, M., and Stefano, G. (1997) *Mol Brain Res* 52, 46-52
38. Salzet, M., Salzet-Raveillon, B., Cocquerelle, C., Verger-Bocquet, M., Pryor, S. C., Rialas, C. M., Laurent, V., and Stefano, G. B. (1997) *J Immunol* 159, 5400-5411
39. Britz, F. C., and Deitmer, J. W. (2002) *Peptides* 23, 2117-2125

40. Britz, F. C., Hirth, I. C., and Deitmer, J. W. (2004) *Eur J Neurosci* 19, 983-992
41. Keating, H. H., and Sahley, C. L. (1996) *J Neurobiol* 30, 374-384
42. Schmidt, J., and Deitmer, J. W. (1999) *Eur J Neurosci* 11, 3125-3133
43. Takahashi, T., Matsushima, O., Morishita, F., Fujimoto, M., Ikeda, T., Minakata, H., and Nomoto, K. (1994) *Zoolog Sci* 11, 33-38
44. Tobin, A. E., and Calabrese, R. L. (2005) *J Neurophysiol* 94, 3938-3950
45. Wang, Y., Price, D. A., and Sahley, C. L. (1998) *Peptides* 19, 487-493
46. Wang, Y., Strong, J. A., and Sahley, C. L. (1999) *J Neurophysiol* 82, 216-225
47. Li, K. W., van Golen, F. A., van Minnen, J., van Veelen, P. A., van der Greef, J., and Geraerts, W. P. (1994) *Mol Brain Res* 25, 355-358
48. Lopez, V., Wickham, L., and Desgroseillers, L. (1993) *DNA Cell Biol* 12, 53-61
49. Santama, N., Wheeler, C. H., Burke, J. F., and Benjamin, P. R. (1994) *J Comp Neurol* 342, 335-351
50. Miller, M. W., Beushausen, S., Vitek, A., Stamm, S., Kupfermann, I., Brosius, J., and Weiss, K. R. (1993) *J Neurosci* 13, 3358-3367
51. Krajniak, K. G., and Greenberg, M. J. (1992) *Comp Biochem Physiol C* 101, 93-100
52. Baratte, B., Gras-Masse, H., Ricart, G., Bulet, P., and Dhainaut-Courtois, N. (1991) Isolation and characterization of authentic Phe-Met-Arg-Phe-NH2 and the novel Phe-Thr-Arg-Phe-NH2 peptide from Nereis diversicolor. *Eur J Biochem* 198, 627-633
53. Krajniak, K. G., and Price, D. A. (1990) Authentic FMRFamide is present in the polychaete Nereis virens. *Peptides* 11, 75-77
54. Diaz-Miranda, L., de Motta, G. E., and Garcia-Arraras, J. E. (1992) *J Exp Zool* 263, 54-67
55. Diaz-Miranda, L., Escalona de Motta, G., and Garcia-Arraras, J. E. (1991) *Cell Tissue Res* 266, 209-217
56. Banvolgyi, T., Barna, J., Csoknya, M., Hamori, J., and Elekes, K. (2000) *Acta Biol Hung* 51, 409-416
57. Gershon, T. R., Baker, M. W., Nitabach, M., Wu, P., and Macagno, E. R. (1998) *J Neurosci* 18, 2991-3002
58. O'Gara, B. A., Brown, P. L., Dlugosch, D., Kandiel, A., Ku, J. W., Geier, J. K., Henggeler, N. C., Abbasi, A., and Kounalakis, N. (1999) *Invert Neurosci* 4, 41-53
59. Calabrese, R. L., Nadim, F., and Olsen, O. H. (1995) *J Neurobiol* 27, 390-402
60. Salzet, M., Bulet, P., Wattez, C., and Malecha, J. (1994) *Eur J Biochem* 221, 269-275
61. Wenning, A., Cahill, M. A., Hoeger, U., and Calabrese, R. L. (1993) *J Exp Biol* 182, 81-96
62. Walker, R. J., Holden-Dye, L., and Franks, C. J. (1993) *Comp Biochem Physiol C* 106, 49-58
63. Norris, B. J., and Calabrese, R. L. (1990) *J Comp Physiol [A]* 167, 211-224
64. Li, C., and Calabrese, R. L. (1987) *J Neurosci* 7, 595-603
65. Kuhlman, J. R., Li, C., and Calabrese, R. L. (1985) *J Neurosci* 5, 2301-2309
66. Kuhlman, J. R., Li, C., and Calabrese, R. L. (1985) *J Neurosci* 5, 2310-2317
67. Salzet, M., Wattez, C., Verger-Bocquet, M., Beauvillain, J. C., and Malecha, J. (1993) *Brain Res* 601, 173-184
68. Evans, B. D., Pohl, J., Kartsonis, N. A., and Calabrese, R. L. (1991) *Peptides* 12, 897-908
69. Wenning, A., and Calabrese, R. L. (1995) *J Exp Biol* 198, 1405-1415
70. Salzet, M., Verger-Bocquet, M., Vandenbulcke, F., and Van Minnen, J. (1997) *Mol Brain Res* 49, 211-221
71. Schaller, H. C. (1976) *Cell Differ* 5, 13-25

72. Schaller, H. C. (1976) A *Cell Differ* 5, 1-11
73. Schaller, H. C., and Bodenmuller, H. (1981) *Proc Natl Acad Sci U S A* 78, 7000-7004
74. Laurent, V., and Salzet, M. (1996) *Regul Pept* 65, 123-131
75. Salzet, M., Vandenbulcke, F., and Verger-Bocquet, M. (1996) *Mol Brain Res* 43, 301-310
76. Matsushima, O., Takahama, H., Ono, Y., Nagahama, T., Morishita, F., Furukawa, Y., Iwakoshi-Ukena, E., Hisada, M., Takuwa-Kuroda, K., and Minakata, H. (2002) *Peptides* 23, 1379-1390
77. Minakata, H., Fujita, T., Kawano, T., Nagahama, T., Oumi, T., Ukena, K., Matsushima, O., Muneoka, Y., and Nomoto, K. (1997) *FEBS Lett* 410, 437-442
78. Minakata, H., Ikeda, T., Nagahama, T., Oumi, T., Ukena, K., Matsushima, O., Kawano, T., and Kimura, Y. (1999) *Biosci Biotechnol Biochem* 63, 443-445
79. Nagahama, T., Ukena, K., Oumi, T., Morishita, F., Furukawa, Y., Matsushima, O., Satake, H., Takuwa, K., Kawano, T., Minakata, H., and Nomoto, K. (1999) *Cell Tissue Res* 297, 155-162
80. Oumi, T., Ukena, K., Matsushima, O., Ikeda, T., Fujita, T., Minakata, H., and Nomoto, K. (1995) *Biochem Biophys Res Commun* 216, 1072-1078
81. Salzet, M., Vieau, D., and Day, R. (2000) *Trends Neurosci* 23, 550-555
82. Kanda, A., Satake, H. , Kawada, T. and Minakata, H. Biochem J 387 (2005) 85-91
83. Kanda, A., Takuwa-Kuroda, K., Iwakoshi-Ukena, E., Furukawa, Y. Matsushima, O. and Minakata, H. J Endocrinol 179 (2003) 281-291
84  R.E. van Kesteren, C.P. Tensen, A.B. Smit, J. van Minnen, P.F. van Soest, K.S. Kits, W. Meyerhof, D. Richter, H. van Heerikhuizen, E. Vreugdenhil and et al., A novel G protein-coupled receptor mediating both vasopressin- and oxytocin-like functions of Lys-conopressin in Lymnaea stagnalis, Neuron 15 (1995) 897-908.

**Figure Legends**

**Figure 1 : Sequence alignment of OT/VP receptor in Invertebrates using Mulatlin software** (http://npsa-pbil.ibcp.fr/cgi-bin/npsa_automat.pl?page=/NPSA/npsa_multalin.html): cephalotocin R and octopressin R [82,83], lysine and arginine-conopressin R [84], annetocin R [4], leech OT/VP R [27]and Hirudotocin R sequence are from leech EST (Salzet, unpublished data). Inset a phylogenetic tree of the Invertebrate receptor using genebee software (http://www.genebee.msu.su/services/phtree_reduced.html).

**Figure 2** : **Sequence alignment of Myomodulin precursor from Mollusks [49,50] and from leech EST** (http://npsa-pbil.ibcp.fr/cgi-bin/npsa_automat.pl?page=/NPSA/npsa_multalin.html):

**Figure 3** : **Sequence alignment of Annelids Excitatory peptide Precursor [76-79]** (http://npsa-pbil.ibcp.fr/cgi-bin/npsa_multalin.html):

**Table 1 : Characterized Annelid Neuropeptides from leeches**

| species | Sequence | Name |
|---|---|---|
| *Theromyzon tessulatum* | SYVMEHFRWDKFGRKIKRRPIKVYPNGAEDE | ACTH-like |
| | SAEAFPLE | Angiotensin I |
| | DRVYIHPFHLLXWG | Angiotensin II |
| | DRVYIHPF | Angiotensin III |
| | RVYIHPF | LORF (leech osmoregulator factor) |
| | IPEPYVWD | |
| | FMRF-amide | FMRF-amide |
| | FM(O)RF-amide | FMRF-amide sulfoxyde |
| | FLRF-amide | FLRF-amide |
| | GDPFLRFamide | GDPFLRF-amide |
| | PLG | MIF-1 |
| | YGGFL | Leucine-enkephalin |
| | YGGFM | Méthionine-enkephalin |
| | YGGFLRKYPK | α neoendorphin |
| | YVMGHFRWDKFamide | MSH-like peptide |
| | GSGVSNGGTEMIQLSHIRERQRYWAQDNLRR | Leech Egg Laying Hormone |
| | RFLEKamide | |
| *Erpobdella octoculata* | DRVYIHPF-amide | Angiotensin II-amide |
| | CFIRNCPKG-amide | Lysine-conopressin |
| | FMRF-amide; FM(O)RF-amide | FMRF-amide |
| | GDPFLRF-amide | GDPFLRF-amide |
| | FLRF-amide | FLRF-amide |
| | IPEPYVWD; IPEPYVWD-amide | LORF |
| *Hirudo medicinalis* | FMRF-amide, FM(O) RF-amide | FMRF-amide |
| | FLRF-amide | FLRF-amide |
| | AMGMLRM-amide, GVSFLRMG, SLDMLRMG, AVSMLRMG, AVSLLRMG | Myomoduline-like peptide |
| | CFIRNCPLG-amide | Hirudotocin |
| *Hirudo nipponia* *Whitmania pigra* | WRLRSDETVRGTRAKCEGEWAIHACLCLGGN amide | Leech excitatory peptide |

**Table 2** : **Characterized Annelids Neuropeptides other Annelids**

| Species | Sequence | Name |
|---|---|---|
| *Perinereis Vancaurica* | AMGMLRMamide | Myomodulin-CARP |
|  | WVGDVQ | Esophagus regulation peptides |
|  | ATWLDT |  |
|  | WMVGDVQ |  |
|  | FYEGDVPY |  |
| *Nereis diversicolor* | FMRF-amide, FM(O)RF-amide | FMRF-amide |
|  | FTRF-amide | FTFR-amide |
| *Nereis virens* | FMRF-amide | FMRF-amide |
|  | FLRF-amide | FLRF-amide |
| *Eisenia Foetida* | CFVRNCPTGamide | Annetocin |
|  | APKCSGRWAIHSCGGNG | GGNG1 |
|  | GKCAGQWAIHACAGGNG | GGNG2 |
|  | RPKCAGRWAIHSCGGGNG | GGNG3 |

**Table 3: Oxytocin/vasopressin peptides discovered in invertebrates**

| Species | Sequence | Name |
|---|---|---|
| *Erpobdella octoculata* | CF**I**RNCP**K**G-amide | Lysine-Conopressin |
| *Hirudo medicinalis* | CF**I**RNCP**L**Gamide | Hirudotocin |
| *Eisenia Foetida* | CF**V**RNCP**T**Gamide | Annetocin |
| *Conus Georgaphus* | CF**I**RNCP**K**G-amide | Lysine-Conopressin |
| *Conus Striatis* | C**I**IRNCP**R**G-amide | Arginine-Conopressin |
| *Octopus* | C**YF**RNCP**I**G-amide | Cephalotocin |
| *Octopus* | CF**W**TSCP**I**G-amide | Octopressin |
| *Locusta migratoria* | C**L**ITNCPRG-amide | Diuretic peptide |